УДК 517.9

# МАТЕМАТИЧЕСКАЯ МОДЕЛЬ СТАБИЛИЗАЦИИ УПРАВЛЕНИЯ РОБОТОМ ИЗ-ЗА ЗАДЕРЖКИ УПРАВЛЯЮЩЕГО СИГНАЛА

Е.В. ЛЕГЧЕКОВА[1], кандидат физико-математических наук, доцент кафедры высшей математики
О.В. ТИТОВ[2], кандидат физико-математических наук, доцент, доцент кафедры естественных наук

[1]*УО «Белорусский торгово-экономический университет потребительской кооперации» г. Гомель, Республика Беларусь*
[2]*УО «Гомельский инженерный институт» МЧС Республики Беларусь, г. Гомель, Республика Беларусь*

Рассматривается вопрос построения системы команд дистанционно управляемого робота, способной осуществлять стабилизацию движения при наличии постоянной задержки управляющего сигнала. Описывается математическая модель робота.

**Ключевые слова:** математическая модель, управляющий сигнал, стабилизация движения, дистанционно управляемый робот.

В настоящее время мобильные роботы являются одной из самых больших областей робототехники. Мобильные роботы находят различное применение в различных задачах, в которых невозможно непосредственное участие человека. Одной из базовых задач управления мобильными роботами является осуществление выхода на заданный уровень линейной и/или угловой скоростей. Данная задача сама по себе довольно простая и хорошо изучена. Однако в ситуациях, когда управляющий сигнал достигает робота не мгновенно, а через некоторое время, качество многих алгоритмов автоматического управления ухудшается, вплоть до потери устойчивости. Запаздывание управляющего сигнала часто возникает, например, при наличии трудоемких вычислений или при удаленном управлении. В работе представлен метод решения описанной проблемы на основе предсказания состояния объекта к моменту действия управляющего сигнала.

**Постановка задачи**

Одной из самых распространенных моделей мобильных роботов является модель с двумя ведущими колесами (рис. 1), управление в которой осуществляется путем задания скоростей правого и левого колес $v_r$ и $v_l$, соответственно. Рассмотрим динамическую модель данного типа роботов [1]:

$$\begin{aligned} m\dot{v}(t) &= F(t) - B_v v(t), \\ J\dot{\omega}(t) &= T(t) - B_\omega \omega(t), \end{aligned} \quad (1)$$

где: $(v, \omega)$ – линейная и угловая скорости робота; $m$ – масса робота; $F$ – равнодействующая сил, действующих на систему; $B_v$ – коэффициент трения поступательного движения; $J$ – момент инерции; $T$ – момент силы; $B_\omega$ – коэффициент трения вращения.

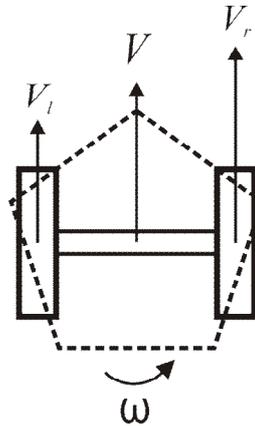

Рис. 1. Модель с двумя ведущими колесами

Силу F и момент силы T можно выразить через силы FR и FL (рис. 2), продуцируемые правым и левым электромоторами робота, соответственно:

$$F(t) = F_R(t) + F_L(t),$$
$$T(t) = l(F_R(t) - F_L(t)), \qquad (2)$$

где: $l$ – расстояние между колесами (или гусеницами) робота.

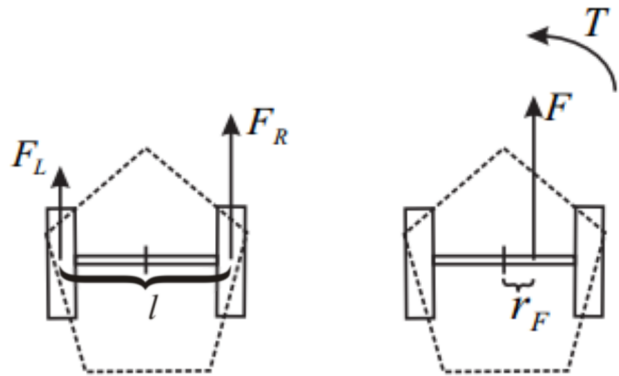

Рис. 2. Динамическая модель робота

Введем новые обозначения: $e_m$ – среднее напряжение на электромоторах, $e_d$ – разность напряжений на электромоторах. Сила $F$ и момент силы $T$ будут пропорциональны этим величинам:

$$F(t) = k_m e_m(t),$$
$$T(t) = k_d e_d(t), \qquad (3)$$

где: $k_m$ и $k_d$ – коэффициенты пропорциональности. Тогда систему (1) можно переписать в следующем виде:

$$\dot{v}(t) = k_v v(t) + k_m e_m(t),$$
$$\dot{\omega}(t) = k_\omega \omega(t) + k_d e_d(t), \qquad (4)$$

Таким образом, динамическая модель робота является линейной стационарной системой. Теперь введем в систему (4) запаздывание управляющего сигнала $h$:

$$\dot{v}(t) = k_v v(t) + k_m e_m(t-h),$$
$$\dot{\omega}(t) = k_\omega \omega(t) + k_d e_d(t-h), \qquad (5)$$

Задача состоит в обеспечении стабилизации системы (5) на заданном уровне $(v^*, \omega^*)$ линейной и угловой скоростей.

**Метод решения**

Рассмотрим систему (4). Введем новые обозначения:

$$x(t) = \begin{vmatrix} v(t) \\ \omega(t) \end{vmatrix}, u(t) = \begin{vmatrix} e_m(t) \\ e_d(t) \end{vmatrix},$$
$$A = \begin{vmatrix} k_v & 0 \\ 0 & k_\omega \end{vmatrix}, B = \begin{vmatrix} k_m & 0 \\ 0 & k_d \end{vmatrix}.$$
(6)

Тогда систему (4) можно переписать в следующем виде:
$$\dot{x}(t) = Ax(t) + Bu(t).$$
(7)

Стабилизируем ее управлением по состоянию:
$$u(t) = Kx(t),$$
(8)

где: $K$ – матрица коэффициентов, обеспечивающая гурвицевость матрицы $A+BK$.

Теперь введем в систему (7) запаздывание управляющего сигнала $h$:
$$\dot{x}(t) = Ax(t) + Bu(t-h),$$
(9)

для нее будем строить управление в виде:
$$u(t-h) = Kx(t),$$
(10)

что можно переписать в таком виде:
$$u(t) = Kx(t+h),$$
(11)

что с очевидностью дает ту же систему, что и без запаздывания. Следовательно, для стабилизации системы (9) необходимо знать прогноз состояния системы на $h$ секунд вперед. Для линейной системы прогноз можно рассчитать при помощи формулы Коши [2]:

$$x(t_2) = e^{A(t_2-t_1)}x(t_1) + e^{A(t_2-t_1)}\int_{t_1}^{t_2} e^{-A(\tau-t_1)}Bu(\tau-h)d\tau.$$
(12)

В частности, полагая $t_1=t$, $t_2=t+h$, получим

$$x(t+h) = e^{Ah}x(t) + e^{Ah}\int_{t}^{t+h} e^{-A(\tau-t)}Bu(\tau-h)d\tau.$$
(13)

Теперь введем обозначение $\theta=\tau-h$. Тогда из (13) имеем:

$$x(t+h) = e^{Ah}x(t) + e^{At}\int_{t-h}^{t} e^{-A\theta}Bu(\theta)d\theta.$$
(14)

Таким образом, управление (11) выглядит следующим образом:

$$u(t) = Ke^{Ah}x(t) + Ke^{At}\int_{t-h}^{t} e^{-A\theta}Bu(\theta)d\theta.$$
(15)

Для реализации управления в виде (15) введем в рассмотрение вспомогательную переменную

$$z(t) = \int_{0}^{t} e^{-A\theta}Bu(\theta)d\theta.$$
(16)

Заметим, что имеет место равенство
$$\int_{t-h}^{t} e^{-A\theta}Bu(\theta)d\theta = \int_{0}^{t} e^{-A\theta}Bu(\theta)d\theta - \int_{0}^{t-h} e^{-A\theta}Bu(\theta)d\theta,$$

или, с учетом (16):
$$\int_{t-h}^{t} e^{-A\theta}Bu(\theta)d\theta = z(t) - z(t-h).$$
(17)

После подстановки (17) в (15) управление принимает вид
$$u(t) = Ke^{Ah}x(t) + Ke^{At}[z(t) - z(t-h)].$$
(18)

Согласно (16), переменная $z(t)$ удовлетворяет дифференциальному уравнению
$$\dot{z}(t) = e^{-At}Bu(t).$$
(19)

Таким образом, компенсирующее управление (11) имеет собственную динамику:

$$\dot{z}(t) = e^{-At}Bu(t),$$
$$u(t) = Ke^{Ah}x(t) + Ke^{At}[z(t) - z(t-h)], \qquad (20)$$

что эквивалентно динамическому регулятору

$$\dot{z}(t) = e^{-At}BKe^{At}z(t) - e^{-At}BKe^{At}z(t-h) + e^{-At}BKe^{Ah}x(t),$$
$$u(t) = Ke^{At}z(t) - Ke^{At}z(t-h) + Ke^{Ah}x(t), \qquad (21)$$
$$z(t) \equiv 0 \forall t \in [-h,0].$$

Данный регулятор позволяет рассчитывать состояние робота через h секунд. Что позволяет отправлять команды роботу необходимые для стабилизации его состояния с учетом этой задержки.

## Литература

**E.V. Legchekova, O.V. Titov**
**MATHEMATICAL MODEL STABILIZATION CONTROL OF A ROBOT DUE TO THE DELAY OF THE CONTROL SIGNAL**

Consider the question of building a system of commands remotely controlled robot that can perform motion stabilization in the presence of a constant delay of the control signal.